\documentstyle[prl,aps,psfig,floats,preprint]{revtex}

\newcommand{\beq}{\begin{equation}}
\newcommand{\eeq}[1]{\label{#1} \end{equation}}

\begin{document}

\title{On the Distribution of Secondaries at High Energies}

\author{L.L. JENKOVSZKY\footnote{E-mail: jenk@gluk.org} and B.V. STRUMINSKY}

\address{Bogolyubov Institute for Theoretical Physics \\
 National Academy of Sciences of Ukraine \\
 Metrologicheskaya str. 14b, 03143 Kiev, Ukraine.}

\maketitle

\begin{abstract}
We show within a geometrical model developed in earlier papers
that multiplicity distributions are cut off at large
multiplicities. The position and motion of the cut-off point is
related to geometrical- and KNO scaling and their violation, in
particular by the rise of the ratio $\sigma_{el}/\sigma_{t}.$ At
the LHC energies a change of the regime, connected with the
transition from shadowing to antishadowing is expected.
\end{abstract}

The properties of multiplicity distribution of secondaries $P(n)$
at high values of $n$ remain among the topical problems in
high-energy physics. As pointed out in a series of recent papers
\cite{Soso}, the underlying dynamics behind these rare processes
may be quite different from the bulk of events.

Our knowledge about high-energy multiplicity distributions comes
from the data collected at the $ISR, \ \ \ Sp\bar pS$ collider
(UA1, UA2 and UA5 experiments) and the Tevatron collider (CDF and
E735 experiments). In should be noticed that the recent results
from the E335 Collaboration taken at the Tevatron \cite{E735} do
not completely agree with those obtained by the UA5 Collaboration
at comparable energies at the $Sp \bar p S$ collider \cite{UA5}
(see Fig. 1).

\begin{figure}[tbh!]
\centerline{
\psfig{figure=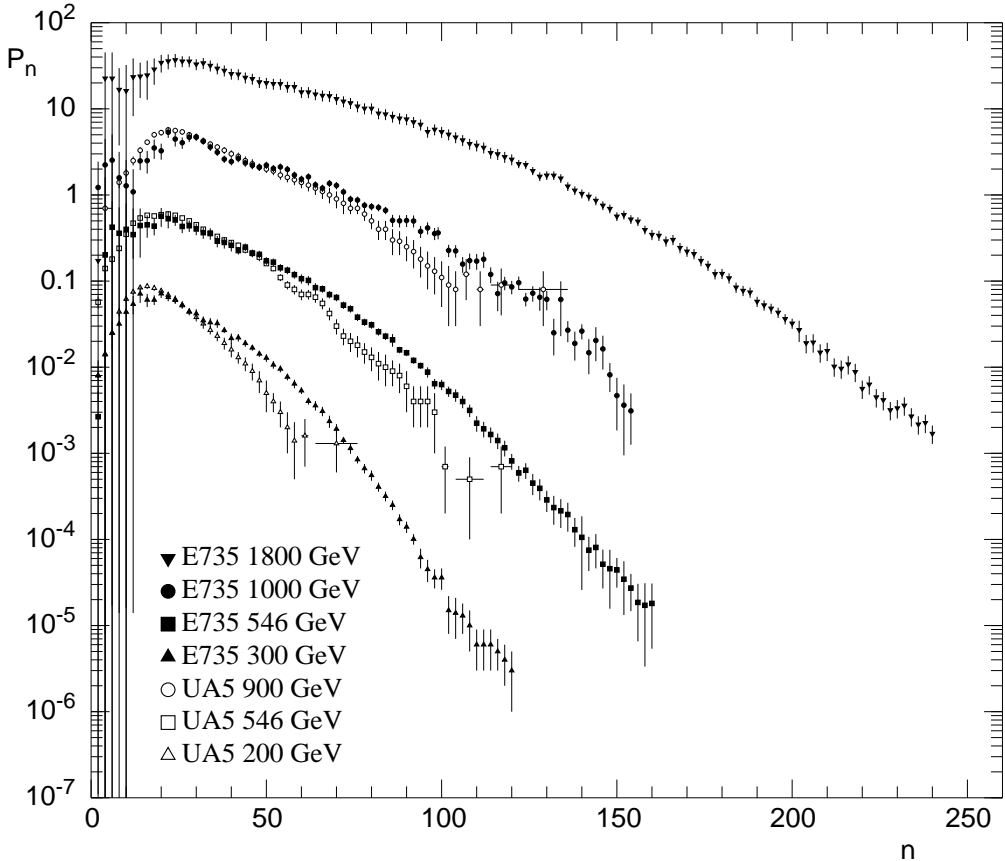,width=0.8\textwidth,angle=0} }
\bigskip
\caption{\small E735 results on charged particle multiplicity
distributions in full phase space compared with UA5 results. Data
from the two experiments which were taken at nearly the same
energy are rescaled by the same factor. } \label{fig}
\end{figure}

 Notice that the delicate features of $\Psi(z)$ at
very large multiplicities, near the large-$z$ edge can be  better
seen if the variable $z$ is used instead of $n.$

 On the theoretical side, it became
common \cite{Dremin}, \cite{GU}, \cite{Kuvshin}, \cite{Odessa} to
approximate the observed distributions by the convolution of two
binomial distributions, accounting for the general "bell-like"
shape of $P(n)$ with the observed structures ("knee" and possible
oscillations) superimposed.

One of the hottest issues in this field is the dynamics of very
high multiplicities (VHM) \cite{Giokaris}, close to the
kinematical limit imposed by the phase space. The VHM events are
very rare, making up only about $10^{-7}$ of the total
cross-sections at the LHC energy, which makes their experimental
identification very difficult. An intriguing question is the
possible existence of a cut-of in the VHM region, beyond
$z=n/<n>\approx 5,$ where $<n>$ is the mean multiplicity. In our
opinion, a better understanding of the underlying physics can be
inferred only in a model involving both elastic and inelastic
scattering, related by unitarity. Such an approach has been
advocated in a series of papers \cite{Aliev}, \cite{Chikovani},
\cite{Kvaratshelia}, summarized in ref. \cite{EChAYa}.

After a brief summary of the main ideas behind this approach, we
analyze the relation of the distribution of secondaries and the
behavior of the elastic and total cross sections with the possible
transition from shadowing to antishadowing \cite{TT}.

We show that the existence of a cut-off at high multiplicities in
the distribution $\Psi(z)$ is related to the validity of GS and
KNO scaling.

The basic idea of the geometrical approach to multiple production,
used in the present paper, is that the number of the secondaries
at a given impact parameter $\rho,$ $n(\rho,s)$  is proportional
to the amount of the hadronic matter in the collision or the the
overlap function $G(\rho, s)$
$$
<n(\rho.s)>=N(s)G(\rho,s), \eqno(1)
$$
where $N(s)$ is related to mean multiplicity, not specified in
this approach, and $G(\rho,s)$ is the overlap function, related by
unitarity to elastic scattering
$$
Im h(\rho,s)=|h(\rho,s)|^2+G(\rho,s).  \eqno(2)
$$
where $h(\rho,s)$ is the elastic amplitude in the impact parameter
representation.
 Unitarity, a key issue in this approach, enters both in the
definition of the elastic amplitude and of the inelastic one (the
overlap function).

In the $u-$ matrix unitarization (see \cite{EChAYa} and references
therein)
$$
G_{in}={\Im u\over{1+2 \Im u+|u|^2}}, \eqno(3)
$$
where $u$ is the elastic amplitude (input, or the "Born term").

 We use a dipole (DP) model for the elastic scattering amplitude,
exibiting geometrical features and fitting the data. After
u-matrix unitarization, the elastic amplitude reads (see
\cite{EChAYa})
$$
h(\rho,s)={u\over{1-i u}},   \eqno(4)
$$
 where $u(y,s)=ige^{-y}, \ \  y={\rho^2\over{4\alpha' L}},$ and $L\equiv\ln s$.

Remarkably, the ratio of the elastic to total cross sections in
this model,
 $$
{\sigma_{el}\over\sigma_{t}}=1-{g\over {(1+g)\ln(1+g)}} \eqno(5)
$$
fixes the (energy-dependent) values of the parameter $g$. Typical
values of $g$ for several representative energies are quoted in
\cite{EChAYa}.

Rescattering corrections to $G_{in}(\rho,s)$ here will be
accounted for phenomenologically according to the following
prescription (see \cite{EChAYa} and earlier reference therein).
$$
G_{in}(\rho,s)=|S(\rho, s)| \tilde G_{in}(\rho,s), \eqno(6)
$$
where $S(\rho,s))$ is the $S$ is the elastic scattering matrix,
related to the $u$ matrix by
$$
S(\rho,s)={1+iu(\rho,s)\over{1-iu(\rho,s)}}. \eqno(7)
$$

This procedure is not unique. For example, it allows the following
generalization (see \cite{EChAYa} and earlier reference therein)
$$
G_{in}(\rho,s)=|S(\rho,s)|^{\alpha}\tilde G^{\alpha}(\rho,s),
\eqno(8)
$$
where $\alpha$ is a parameter, varying between 0 and 1.

We assume
$$
<n(\rho,s)>=N(s)\tilde G^{\alpha}_{in}(\rho,s). \eqno(9)
$$

The moments are defined by (see \cite{EChAYa} and earlier
references therein)
$$
<n^k(s)>={N^k(s)\int G_{in}(\rho,s)(G_{in}^{alpha}(\rho,s))^k
d^2\rho\over{\int G_{in}(\rho,s)d^2\rho}} \eqno(10)
$$
Now we insert  the expression for the DP with the u-matrix
unitarization (4) into (10) to get
$$
<n^k(s)>={N^k(s)(1+g)\over
g}\int_0^g{dx\over{(1+x)^2}}\Biggl(\Bigl({1+x\over{1-x}}\Bigr)^{\alpha}{x\over{(1+x)^2}}\Biggr)^k.
\eqno (11)
$$
 The mean multiplicity $<n(s)>$ is defined as
 $$
 <n(s)>={N(s)(1+g)\over
 g}\int_0^g{x dx\over{(1+x)^4}}\Bigl({1+x\over{1-x}}\Bigr)^{\alpha}={N(s)\over a}. \eqno (12)
$$

For the distributions we have
$$
P(n)={1+g\over
g}\int_0^g{dx\over{(1+x)^2}}\delta\Biggl(n-N\Bigl({1+x\over{1-x}}\Bigr)^{\alpha}{x\over{(1+
x)^2}}\Biggr).
\eqno (13)
$$

Integration in (13) gives
$$
\psi(z)=<n>P(n)={1+g\over g}{x(1-x)\over{z(1+x)[(1-x)^2+2\alpha
x}]},
$$
where $z=n/<n>.$

 Since the above integral is non-zero only when the
argument of the $\delta$ function vanishes,
$$
n=N\Biggl({1+x\over{1-x}}\Biggr)^{\alpha}{x\over{(1+x)^2}},
$$
one gets a remarkable relation
$$
z={ax \over (1+x)^{2-\alpha}(1-x)^{\alpha}}. \eqno(14)
$$

To calculate the distribution $\Psi(z)$ one needs the solution of
equation (14). It can be found explicitly for two extreme cases,
namely $\alpha=0$ and $\alpha=1.$ Otherwise, it can be calculated
numerically.

The maximal value of $z$, corresponding to $x=g$ ($x$ varies
between $0$ and $g$), can be found as:
$$
z_{max}={ag\over{(1+g)^{2-\alpha}}|(1-g)|^{\alpha}}. \eqno(15)
$$

It can be seen from (15) that $z_{max}$ is a constant if $g$ is
energy independent. The experimentally observed ratio
$\sigma_{el}/\sigma_t$ varies between 53 Gev and 900 Gev from
0.174 to 0.225, implying the variation of g from 0.489 to 0.702,
uniquely determined by the above ratio. This monotonic increase of
g(s) in its turn pushes $z_{max}(s)$ outwards, terminating when
$g$ reaches unity (according to \cite{EChAYa} this will happen
around 10 TeV, i.e. at the future LHC), whenafter the term $|1-g|$
in (15) will start rising again, pulling $z_{max}(s)$ back to
smaller values. I.e. $z_{max}(s)$  has its own maximum in $s$ at
$g=1.$

The unusual behavior of $z_{max}(s)$ is not the only interesting
feature of the present approach. This effect can be related to the
behavior of the ratio $\sigma_{el}/\sigma_t$.  As argued by
Troshin and Tyurin (see \cite{TT}), $\sigma_{el}/\sigma_t$
 may pass the so-called black disc limit and
continue rising in a new, "antishadowing" mode of the $u$-matrix
unitarity approach (multiplicity distributions were not considered
in that paper). According to the recent calculations \cite{DJS}
the transition from shadowing to antishadowing will also occur in
the LHC energy region.

To summarize, we found a regularity connecting the geometrical
properties in high-energy dynamics (GS and KNO scaling) with the
dynamics of the high-multiplicity processes. We showed, in
particular, that exact geometrical, or KNO scaling, implying
constant $g$ in our model, results in a cut-off at large $z$ of
the distribution function $\Psi(z).$ Any departure from scaling
(energy dependence of $g$ in our model) shifts the point $z_m$
according to eq. (15). Within the present accelerator energy
domain (ISR, SPS, Tevatron) $g$ varies from about 0.5 to about
0.8. It will reach the critical value $g=1$ at LHC, where we
predict a change of the regime: $z_{max}(s)$ will start decreasing
and the black disc limit will be passed (which, as shown in
\cite{TT} and \cite{DJS}, is not equivalent to the violation of
the unitarity limit, but means passage from shadowing to
antishadowing \cite{TT} and \cite{DJS}).

Finally, it should be noted that we use many model assumptions,
decreasing the predictive power of our calculations. These
assumptions concern mostly the way absorption corrections are
introduced and the assumption of the local ($\delta$ function)
dependence of multiplicities on the impact parameter. Both
assumptions, as well as others can be modified. As a result we
quantitive rather than qualitative changes in the results.

\vfill \eject

\end{document}